\documentclass[aps,prl,twocolumn,showpacs,amsmath,amssymb,letterpaper, floatfix, 10pt,superscriptaddress]{revtex4-1}
\usepackage[pdftex]{graphicx} 
\usepackage{dcolumn} 
\usepackage{bm} 
\usepackage{relsize}
\usepackage{amsmath}
\usepackage{amssymb}
\usepackage{color}
\usepackage[
pdfstartview=XYZ,
bookmarks=false,
colorlinks=true,
linkcolor=blue,
urlcolor=blue,
citecolor=blue,
pdftex,
bookmarks=false,
linktocpage=true,
hyperindex=false
]{hyperref}
\usepackage{psfrag}
\usepackage[integrals]{wasysym}
\usepackage{epstopdf}
\usepackage{tocloft}
\usepackage{pdfpages}

\newcommand{\mycomment}[1]{}

\begin{document}
\title{Spin relaxation via exchange with donor impurity-bound electrons}
\author{Lan Qing}
\affiliation{Department of Physics and Astronomy, University of Rochester, Rochester, NY, 14627}
\affiliation{Department of Physics, Center for Nanophysics and Advanced Materials, U. Maryland, College Park, MD 20742}
\author{Jing Li}
\author{Ian Appelbaum}
\altaffiliation{appelbaum@physics.umd.edu}
\affiliation{Department of Physics, Center for Nanophysics and Advanced Materials, U. Maryland, College Park, MD 20742}
\author{Hanan Dery}
\affiliation{Department of Physics and Astronomy, University of Rochester, Rochester, NY, 14627}
\affiliation{Department of Electrical and Computer Engineering, University of Rochester, Rochester, NY, 14627}

\begin{abstract}
At low temperatures, electrons in semiconductors are bound to shallow donor impurity ions, neutralizing their charge in equilibrium. Inelastic scattering of other externally-injected conduction electrons accelerated by electric fields can excite transitions within the manifold of these localized states. Promotion of the bound electron into highly spin-orbit-mixed excited states drives a strong spin relaxation of the conduction electrons via exchange interactions, reminiscent of the Bir-Aronov-Pikus process where exchange occurs with valence band hole states. Through low-temperature experiments with silicon spin transport devices and complementary theory, we reveal the consequences of this previously unknown spin depolarization mechanism both below and above the impact ionization threshold.

\end{abstract}
\maketitle

Spin exchange is central to many physical mechanisms that drive interactions between internal degrees of freedom in otherwise decoupled systems. In condensed-matter physics, exchange arises in diverse examples such as the well-known Overhauser and Knight effects between electron and nuclear spins \cite{Overhauser_PR53,Knight_PR49,Solomon_PR55, Chan_PRB09}, Glauber or Kawasaki kinetics in Ising models of ferromagnetism \cite{Hohenberg_RMP77}, and in the many-body RKKY and Kondo effects \cite{Bayat_PRL12,Kondo_JPSJ05}. Within atomic physics, it is essential in optical pumping of noble gas nuclei for subsequent spin resonance detection \cite{Walker_RMP97, Happer_RMP72}. The generality of this phenomenon extends even to the realm of particle physics, as in pion-nucleon isospin-exchange scattering at the $\Delta$-resonance \cite{Lindenbaum_ARNS57, Gell-Mann_PR62}. Direct impact on computing technology may one day occur as well, if robust qubits can be constructed from the spin of electrons bound to shallow donor impurity potentials in group-IV elemental semiconductors \cite{Kane_Nature98}. Spin exchange can then play an especially important role as the physical basis for state initialization and entanglement \cite{Loss_PRA98, Petta_Science05, Anderlini_Nature07, Amico_RMP08}.

In this Rapid Communication, we demonstrate experimentally and describe theoretically how inelastic scattering between conduction and impurity-bound electrons leads to strong depolarization of both spins via mutual spin exchange. As we show with low-temperature spin transport measurements in unintentionally-doped silicon devices, this mechanism far outweighs the otherwise-dominant Elliott-Yafet spin relaxation mechanism \cite{Yafet_SSP63,Elliott_PR54}. The latter is mediated by the weak spin-orbit mixing of Pauli states in the conduction band of non-degenerate silicon \cite{Li_PRL11,Song_PRB12,Cheng_PRL10,Tang_PRB12, Jansen_NatureMater12}. Incorporating the exchange in a master equation approach successfully reproduces the observed nonlinear dependence of the charge and spin currents on temperature and electric field. In addition, this work includes a detailed formalism of the spin-dependent transitions in impurity states that may elucidate the physics relevant to terahertz laser emission from shallow donors in silicon \cite{Pavlov_PRL00}.

\begin{figure*}[t!]
\center\includegraphics[width=17cm]{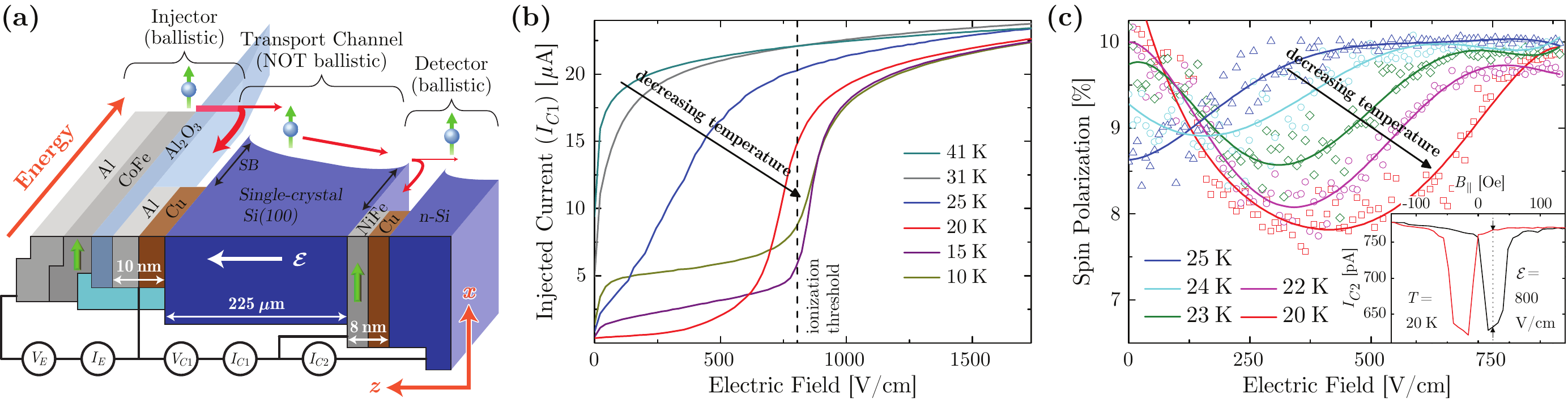}
\caption{(a) Schematic side view of the spin transport device. Energy band diagram is shown as depth into the page. Temperature dependence of (b) injected electron current, and (c) spin polarization from transport through a 225~$\mu$m-thick unintentionally doped Si device with approximately $10^{12}$~cm$^{-3}$ donor concentration (the lines merely guide the eye). The hot electron injector tunnel junction voltage used here is $V_E=-1.3$~V, resulting in approximately 30~mA tunnel current that is unaffected by temperature or electric field in the Si transport region. The inset to (c) shows a characteristic spin-valve measurement from which the spin polarization can be determined (the magnetic field $B_\parallel$ is directed in the plane of the magnetic films).}
\label{fig:expr}
\end{figure*}

As schematically shown in Fig.~\ref{fig:expr}(a), our experiments utilize a ferromagnetic thin-film cathode to perform tunnel-junction injection of \emph{spin-polarized} hot electrons into unintentionally very low-doped \textit{n}-type Si(100): $L=225$~micron-thick float-zone grown wafer with room-temperature resistivity $\rho\approx5$~k$\Omega\cdot$cm \cite{Huang_APL07}. This resistivity is two orders of magnitude less than that of intrinsic silicon, indicating the presence of highly-soluble phosphorus donors in our device at a level of approximately $N_d=10^{12}$~cm$^{-3}$ \cite{Thomas_JCG90}.

Figure~\ref{fig:expr}(b) shows that for temperatures down to $\approx25$~K, the measured injection current ($I_{C1}$) of spin-polarized electrons is determined essentially only by the tunnel junction emitter voltage $V_E$ (when $qV_E$ exceeds the otherwise-rectifying Schottky barrier height of $\approx0.7$~eV at the metal thin-film contact interface). Furthermore, using a spin detector based on ballistic hot electron transport through a ferromagnetic thin film \cite{Appelbaum_Nature07, Lu_APL10}, we can measure the relative difference in spin transport signal between parallel and antiparallel injector/detector magnetic configurations as shown by sample data in the inset to Fig.~\ref{fig:expr}(c). The spin polarization $P$ derived from it rises and saturates as a function of internal electric field ($V_{C1}/L$) for 25~K and higher temperatures, as shown in Fig.~\ref{fig:expr}(c). This behavior is due to the increase in drift velocity $v$ and reduction in transit time through the Si transport region \cite{Kameno_APL12}, from which a temperature-dependent spin lifetime $\tau$ can be determined in conjunction with spin precession measurements via  $P\propto\exp(-\frac{L}{v\tau})$ \cite{Huang_PRL07, Huang_PRB10}.

However, for lower temperatures this conventional behavior changes. Our central experimental result and the primary focus of this Rapid Communication is shown in Fig. ~\ref{fig:expr}(c). We find a stark change in the detected spin polarization when the temperature drops below 25~K: the spin polarization initially \emph{decreases} with increasing electric field and eventually recovers to larger values at strong fields. We will find that this intriguing non-monotonic dependence is a result of inelastic spin-exchange scattering between conduction and localized electrons in the bulk silicon.

At these reduced temperatures, charge transport data in Fig.~\ref{fig:expr}(b) shows that the injected current is rapidly suppressed for internal electric field $\mathcal{E}\apprle800$~V/cm, despite a constant incident flux of ballistic hot electrons impinging on the injection interface from a steady emitter voltage $V_E$. This transition temperature roughly corresponds to thermalization of conduction electrons into the $E_{D^0}\approx 45$~meV donor state to form the $D^0$ neutral ground state below $\approx E_{D^0}/(k_B\ln{N_c/N_d})\approx30$~K, where $N_c$ is the conduction band effective density of states. With the restoration of charge transport for  $\mathcal{E}\apprge800$~V/cm, the conduction electron spin polarization also recovers. Due to reduction in signal current from freeze-out effects in the detector portion of our device, spin polarization after transport can be measured only down to $\approx20$~K (i.e. $I_{C2}\rightarrow 0$ for $T<20$~K).

\begin{figure}[b!]
\center\includegraphics[width=6.5cm]{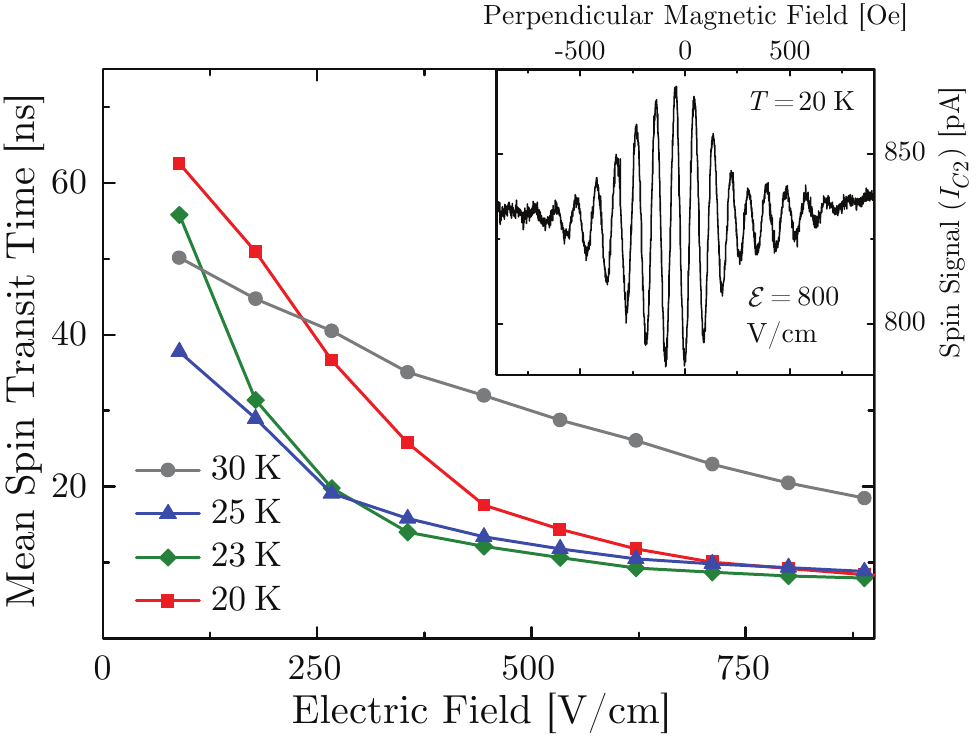}
\caption{Spin transit time through nominally undoped Si. Inset shows characteristic spin precession oscillations from which transit times are determined via Fourier transform at $\mathcal{E}=800$~V/cm and $T=20$~K.
}
\label{fig:hanle}
\end{figure}

To probe the electron transport in greater detail, spin precession measurements performed in an out-of-plane magnetic field can be used to determine the time-of-flight of electrons traveling through the Si channel with a transform method \cite{Huang_PRB10}. Results from these measurements are shown in Fig.~\ref{fig:hanle}.  As temperature decreases from 30~K, mean transit times also initially decrease; this reduction is consistent with the expected higher mobility due to suppression in electron-phonon scattering. However, for temperatures below $\approx25$~K, transit times begin to \emph{increase} in the same electric field region $\mathcal{E}\apprle800$~V/cm where the charge current is suppressed. This behavior implies the role of transient interactions with impurity potentials in the electron transport \cite{Lu_PRL11}.

\begin{figure*}[t!]
\center\includegraphics[width=17cm]{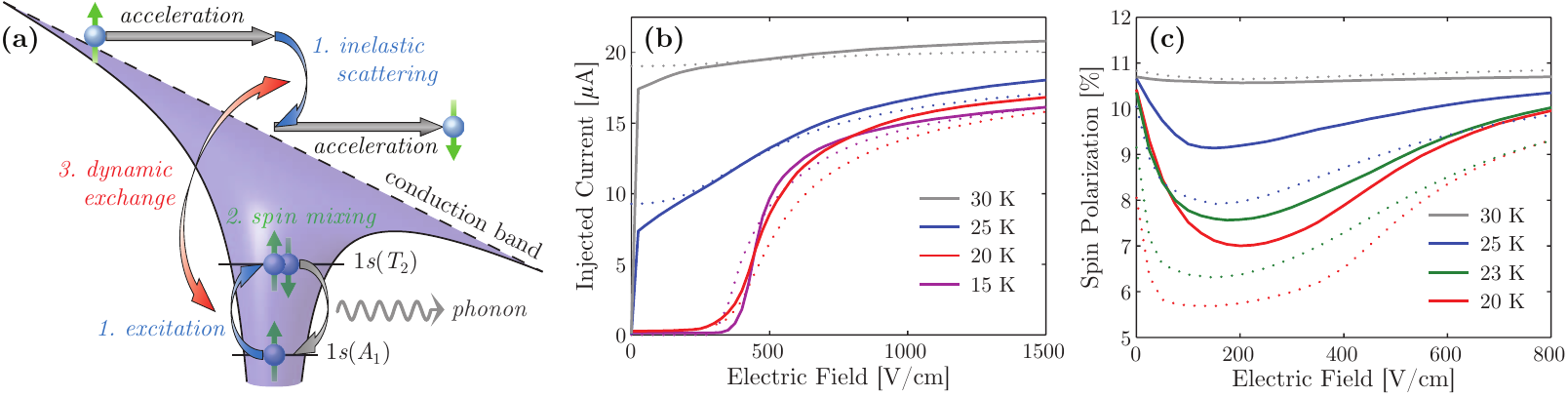}
\caption{(a) Simplified schematic of exchange-driven spin relaxation with neutral donors. In the first step, inelastic scattering with an accelerated conduction electron excites the bound electron ($A_1 \rightarrow T_2$). In the second step, ultrafast spin relaxation of the bound electron is facilitated by spin-mixing of the excited state ($T_2$). In the third step, the conduction-electron spin suffers relaxation via exchange. The electric field is exaggerated for illustrative purposes. Calculated (b) current and (c) spin polarization from transport through the silicon channel as a function of electric field. Solid lines are obtained from solution of the master equations [Eq.~(\ref{eq:rate})], and dotted lines from the simplified model [Eqs.~(\ref{eq:ion_simp})-(\ref{eq:spin_simp2})]. These results reproduce the nonlinear dependencies on temperature and electric fields as measured in the experiment [Figs. \ref{fig:expr}(b)-(c)].}
\label{fig:model}
\end{figure*}

Our physical picture used to explain this observed phenomena relies on inelastic scattering of energetic conduction band electrons with those bound to localized impurity potentials. ``Impact excitation" occurs when the scattering event results in an excited but still bound state, and ``impact ionization" when the donor-bound electron is liberated into the continuum of the conduction band \cite{Mitin_PRB86}. The ionization rate is a strong function of the accelerating electric field; once it is comparable to the recombination rate (at the so-called ``breakdown field'' \cite{Kaiser_PRL59}) the free carrier concentration rises abruptly due to a chain reaction, similar to the multiplication process in ``avalanche'' photodetectors. As a result, nearly all donors are ionized above the breakdown field at any temperature.

The existence of impurity levels more weakly bound than the ground state can reduce this breakdown field ionization threshold. For example, once the localized electron is excited to the $2p_0$ state, it is more likely to undergo thermal activation to the conduction band \cite{Pavlov_PRL00}. The resulting $< 1$~kV/cm scale -- lower than the regime needed to ignite significant intervalley $f$-process phonon scattering \cite{Li_PRL12} -- agrees well with the observed ionization threshold in Fig.~\ref{fig:expr}(b).

Donor states in silicon are formed from the sixfold valley degeneracy of the conduction band; the 1$s$ hydrogenic states are split by the crystal field (`valley-orbit interaction') into a nondegenerate  $1s(A_1)$ ground state, a twofold-degenerate $1s(E)$ level, and a threefold-degenerate $1s(T_2)$ level \cite{Kohn_BC57}.  Because of spin-orbit interaction, the latter state in particular becomes highly spin mixed $\Gamma_4\rightarrow\Gamma_8\oplus\Gamma_7$ \cite{Song_PRL14,Castner_PR67,Kohn_BC57}, in a way analogous to the $p$-like light and split-off hole states in the valence band of cubic semiconductors. Therefore, impact excitation from $A_1$ to $T_2$, accompanied by $\sim$10~meV energy loss from the conduction electron \cite{Castner_PR67}, creates a superposition of stationary bound states whose spin precesses during time evolution. Upon subsequent stochastic phonon emission and return to the $A_1$ level, this spin is efficiently depolarized with respect to its initial orientation.  As illustrated in Fig.~\ref{fig:model}(a), simultaneous exchange couples the conduction electron to this spin loss mechanism \cite{Ando_arxiv14}. Similarly, spin depolarization of conduction electrons can be mediated via exchange between the unpolarized bound electron after its return to the $A_1$ state and a newly injected polarized electron that arrives at the impurity vicinity at later times \cite{Lo_PRL11, Ghosh_PRB92}, similar to conduction electron spin exchange with valence band holes in the well-known Bir-Aronov-Pikus relaxation mechanism \cite{Bir_ZETF75,*Bir_SPJETP75}.

We can thus qualitatively understand the trends in Fig.~\ref{fig:expr}(c): Increasing electric field drives impact excitation and its concomitant conduction electron depolarization. This process continues until impact ionization at and above the breakdown field eliminates the population of bound electrons necessary for the process to occur. Spin polarization therefore is restored.

In an initial numerical model incorporating this phenomenology, we consider only a single effective impurity state $i$ (i.e., ignoring their fine-structure, valley-orbit splitting, and different discrete levels) with depolarization caused by round-trip transitions to a spin-mixed state $i'$. Our spin injector sources spin density at a rate $P_Jn_cR$, where $P_J$ is initial spin polarization at injection ($11.5\%$ to match the experiment), $n_c$ is the conduction electron number density, and $R$ is the injection rate. The latter can be empirically estimated from injection current density ($\approx 10^{-1}$~A/cm$^2$) and the transport length as $\approx10^8$~s$^{-1}$. 

Spin relaxation due to the exchange-driven mechanism described here results in a measured conduction electron spin polarization $P_c$ that is necessarily less than the injection polarization $P_J$. Conservation of angular momentum requires that the conduction electron spin density lost to the bound electrons per unit time $\left[ n_cR(P_J-P_c) \right]$ is equal to the quantity gained by the impurity electron via elastic exchange $\left[ \chi n_c n_i(P_c-P_i) \right]$, where $\chi$ is the elastic exchange rate coefficient, and $n_i$($P_i$) is the impurity-bound electron density(spin polarization). In steady state, this same quantity is lost by the impurity to the environment $\left[P_i(\omega_{i,i'}n_i+\gamma_{i,i'} n_c n_i)\right]$, where the first and last terms account for electron-phonon Castner-Orbach spin depolarization \cite{Castner_PR67} with virtual transitions to a spin-mixed state $i'$ (proportional to temperature-dependent rate coefficient $\omega_{i,i'}$) and inelastic exchange via impact excitation (proportional to the strongly electric-field-dependent coefficient $\gamma_{i,i'}$), respectively. Thus, we have the system of coupled equations 

\begin{align}
 n_cR(P_J-P_c) &= \chi n_c n_i(P_c-P_i) \notag \\
 &= P_i(\omega_{\uparrow\downarrow}n_i+\gamma_{i,i'} n_c n_i).\label{eq:ion_simp}
\end{align}

Along with the steady-state rate equation for conduction electron density
\begin{equation}
\omega_{i,c} n_i - \omega_{c,i} n_c+\gamma_{i,c} n_c n_i =\frac{dn_c}{dt}=0,\label{eq:spin_simp2}
\end{equation}
and local charge neutrality $n_i=N_d-n_c$, we can algebraically solve for both the conduction electron density which determines the measured current and the conduction electron polarization sensed by our spin detector. Here, $\omega_{i,c}$ is the  phonon-assisted thermal (Arrhenius) transition rate into the conduction band, $\omega_{c,i}$ is the static thermalization rate, and $\gamma_{i,c}$ is the electric field-dependent impact ionization rate. Note that when the latter vanishes in zero-field equilibrium, this equation yields thermodynamic detailed balance.

The numerical results using appropriate coefficients obtained by comparison to experiment [e.g. Ref. \onlinecite{Castner_PR67}] and Monte-Carlo simulations [e.g. Ref. \onlinecite{Li_PRL12}], which should be compared to the corresponding experimentally measured values in Fig.~1(b)-(c), are shown by dotted lines in Fig.~2(b)-(c). Even with this inexact, minimal model, they already reconcile the main trends in the empirical observations.

To remove the phenomenological nature of the simplified approach above, we can incorporate all the relevant processes with a system of general master rate equations for the occupations of the $\ell$th valley-orbit level with spin $\sigma$
\begin{align}\label{eq:rate}
\frac{\partial n_{\ell\sigma}}{\partial t}=&\;G_{\ell\sigma}+\overbrace{\sum_{\ell',\sigma'}\left(\omega_{\ell'\sigma'\!,\ell\sigma}n_{\ell'\sigma'}-\omega_{\ell\sigma,\ell'\sigma'}n_{\ell\sigma}\right)}^\text{phonon absorption/emission}\\
&+\overbrace{\sum_{\ell'\neq\ell,\sigma'\neq\sigma}\chi_{\ell,\ell'}\left(n_{\ell\sigma'}n_{\ell'\sigma}-n_{\ell\sigma}n_{\ell'\sigma'}\right)}^\text{elastic exchange}\nonumber\\
&+\underbrace{\sum_{\ell',\sigma',\sigma''}\left(\gamma^{\sigma''}_{\ell'\sigma'\!,\ell\sigma}n_{\ell'\sigma'}n_{c\sigma''}-\gamma^{\sigma''}_{\ell\sigma,\ell'\sigma'}n_{\ell\sigma}n_{c\sigma''}\right)}_\text{inelastic exchange + impact},\nonumber
\end{align}
where we mark the separate terms with their physical meanings. Conduction electrons are denoted by $\ell =c$, and localized electrons by $\ell=0,1,2\ldots$ including all possible $1s$, $2s$, and $2p_0$ states with each comprised of twelve components (2 for spin and 6 for valley). $G_{\ell\sigma}$ denotes the conduction electron spin density lost to the bound electrons per unit time. As in the simple phenomenological model, the $\omega$ coefficients are phonon-assisted thermal transition rates, $\chi$ coefficients are the exchange rates (per unit density), and $\gamma$ parameters are the electric field-dependent impact ionization or excitation rates (per unit density). Subscripts stand for the corresponding initial and final states, and superscripts of $\gamma$ indicate the spins of impact conduction electrons, which accounts for the different nature of singlet and triplet scattering \cite{Honig_PRL66}. In this rigorous approach, all these rates are calculated directly from Fermi's golden rule to first order, except orbital-conserving spin-flip transitions necessarily involving two-phonon processes with virtual states, which are treated in second order \cite{Castner_PR67}. The detailed calculations are given in the supplemental material \cite{Note1}, where we use Monte Carlo simulation to generate the energy distribution of electric field-heated conduction electrons \cite{Li_PRL12,Jacoboni_RMP83}, and include spatial dependence of $n_{\ell\sigma}$ by an exponential relation from the simple drift-diffusion model. We perform explicit time-domain simulation starting from fully ionized initial conditions to obtain the steady-state solution without any assumption of absolute charge neutrality. The results of this exact calculation for different electric field and temperature are summarized by the solid lines in Fig.~\ref{fig:model}(b)-(c) and agree with the experimental results shown in in Fig.~\ref{fig:expr}(b)-(c).

The exchange spin relaxation mechanism shown by our experiment [Fig.~\ref{fig:expr}(c)] and theory [Fig.~\ref{fig:model}(c)] outweighs the electron-phonon contribution \cite{Li_PRL11} at low temperatures. Furthermore, it exists even without the accelerating electric field, in thermal equilibrium when it occurs due to energy loss by conduction electrons initially in the Maxwellian tail. As measured by L{\'e}pine, at low temperatures most of the donors are occupied by electrons and the conduction electron spin lifetime exhibits an anomalous increase with temperature \cite{Lepine_PRB70}. The spin relaxation mechanism described here can easily account for this behavior: upon increase in temperature, the drop in neutral donor density outweighs enhancement of donor spin relaxation from higher thermal energy. The conduction electron spin flip rate associated with this exchange mechanism then falls with temperature until it competes with electron-phonon Elliott-Yafet spin relaxation to determine the total spin lifetime.

We have performed complementary experiments that exclude possible contributions to the observed phenomena from interface effects at the metal-semiconductor junction \cite{Lu_APL13} as well as from electron-electron scattering in the conduction band \cite{Dimitrova_JETPL2008}.  As shown in Sec. 1 of the supplemental material, the exchange spin relaxation mechanism is not observed when we use shorter undoped silicon channels (10 $\mu$m), showing that interface effects are irrelevant. Furthermore, this effect is not due to the presence of the two-electron charged donor $D^-$ \cite{Lu_APL14}, because these weakly-bound states form only at much lower temperatures \cite{Thornton_PRL73}. The effect is most clearly observed when the densities of conduction and localized electrons are comparable, and when the channel is long enough such that injected electrons cannot avoid interacting with impurities on their way to the detector.

In closing, we notice that stimulated emission from the relevant $2p_0\rightarrow1s(E)$ transition in phosphorus doped silicon can be used for the realization of a terahertz laser \cite{Pavlov_PRL00}. The master equations presented in this Rapid Communication include both these states,  so, along with spin-dependent radiative dipole selection rules, spin-polarized carrier injection may be shown to allow external control over circular polarization of the output terahertz electromagnetic field; in this case, alternative spin injection or generation schemes may be required such as interband optical orientation \cite{Lampel_PRL68, Sircar_PRB14, Li_PRB10, Li_PRB13}. Meanwhile, since spin-polarized donor-bound electrons are integral to Kane's proposal for a phosphorus nuclear spin-based silicon quantum computing architecture \cite{Kane_Nature98}, it is hoped that the picture unraveled in this work will yield insight relevant to the robustness of solid-state implementations of quantum information.

We acknowledge helpful comments by Dr.~Yang~Song. Work at UMD was supported by the Office of Naval Research under contract N000141410317, the National Science Foundation under contract ECCS-1231855, the Defense Threat Reduction Agency under contract HDTRA1-13-1-0013, and the Maryland NanoCenter and its FabLab. Work at UR was supported by the National Science Foundation under contract ECCS-1231570 and the Defense Threat Reduction Agency under contract HDTRA1-13-1-0013.


%

\clearpage


\section*{Supplementary material}

Lan Qing,[1,2] Jing Li,[2] Ian Appelbaum,[2,*] and Hanan Dery[1,3]

[1]*Department of Physics and Astronomy, University of Rochester, Rochester, NY, 14627*
[2]*Department of Physics, Center for Nanophysics and Advanced Materials, U. Maryland, College Park, MD 20742*
[3]*Department of Electrical and Computer Engineering, University of Rochester, Rochester, NY, 14627*

**1. Additional experimental data.** To further establish the extrinsic nature of the measured effects, we have varied the injected current $I_{C1}$ over a factor of $\approx 5$ by changing the tunnel junction voltage $-1.1$ V $< V_E <$ $-1.5$ V, a range limited by signal to noise and tunnel junction stability. Although the charge transport current dependence on voltage [shown in the inset to Fig. I(a)] is only slightly modified besides overall magnitude scaling, the transit-time delay shown in Fig. I(a) is enhanced by weaker injection and the spin polarization spectroscopy shown in Fig. I(b) is greatly affected. This behavior can be understood by noting that when we saturate the system with higher injected electron density, the relative contribution from impurity depolarization is diluted due to the fixed impurity concentration; the spin depolarization effect is therefore weakened. A lower injected density enhances the depolarization because a greater proportion of conduction electrons can interact with the finite number of impurities.

On devices of shorter transit length but comparable impurity density, the signal suppression is absent, as shown in Fig. II(a). The suppression is therefore a bulk effect rather than one controlled by scattering at the injection interface. However, a similar suppression is seen in short-channel devices with doping in the $10^{14}$ cm$^{-3}$ range as demonstrated in Fig. II(b), making the total number of impurities roughly equal to the nominally undoped 225-$\mu$m devices. This behavior can be understood from the effective interaction length of conduction and localized electrons, which is scaled by the ratio of their densities. This quantity should be comparable to the device length for the impact effects to be clearly observed.

**2. Impurity states.** To formulate the impact and ex-

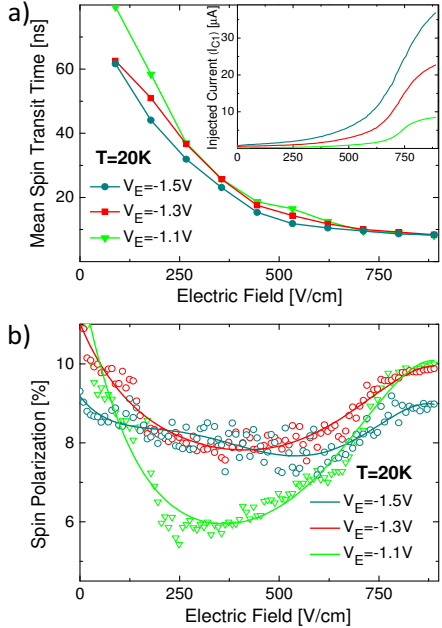

FIG. I. Effect of injected current variation on (a) spin transit time and (b) electric-field dependent spin depolarization [lines merely to guide the eye] for injected conduction electrons in unintentionally-doped $n$-Si at 20 K. The enhancement of depolarization under more dilute injection shown here is consistent with the theory of dynamic spin exchange.

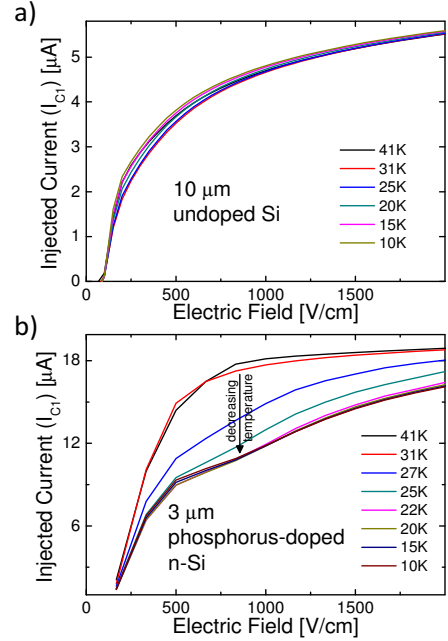

FIG. II. Results from different devices: (a) Injected current at low electric fields is unaffected by low temperatures in similarly unintentionally-doped devices with much shorter transport distance (10-$\mu$m). (b) When intentionally doped in the range $10^{14}$ cm$^{-3}$, the suppression is evident even in these 3-$\mu$m-thick devices.

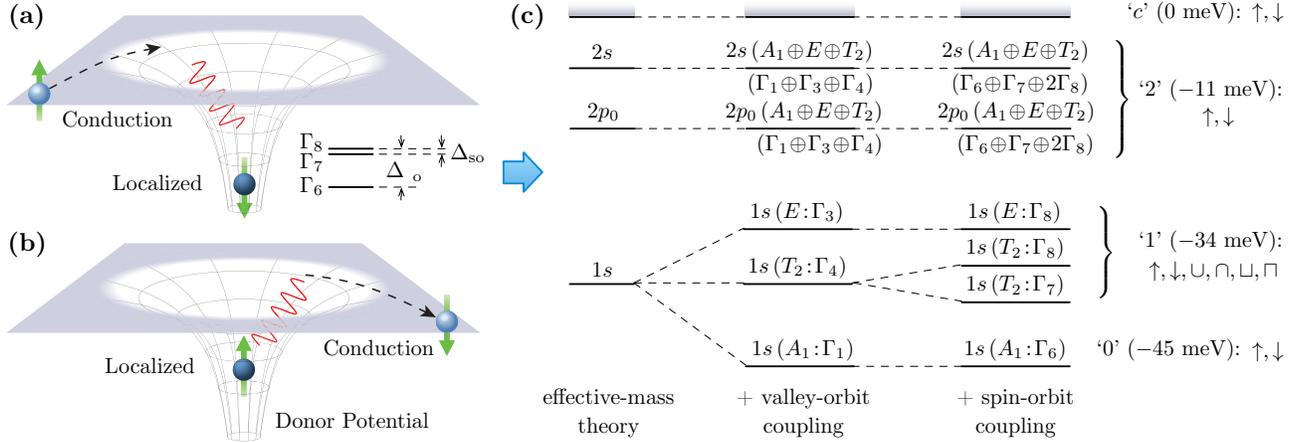

FIG. III. Illustrations (a) before and (b) after the inelastic exchange scattering between conduction and localized electrons. The twelve $1s$ donor states split into those with $\Gamma_6$, $\Gamma_7$ and $\Gamma_8$ symmetries. The splitting energies are determined by the valley-orbit coupling $\Delta_{\text{vo}}$ and the spin-orbit coupling $\Delta_{\text{so}}$. (c) Detailed scheme of the donor energy levels in silicon (not to scale). The letters in parentheses indicate the irreducible representations of $T_d$, with the notation in Ref. [S1] (see text).

change effects quantified by our model and depicted by Fig. III(a)-(b), we need to analyze the impurity states. For phosphorus-doped silicon, it is sufficient to consider only the $1s$, $2s$, and $2p_0$ states [S2], since the upper states are close to each other and approximately form a continuum which extends to the conduction band. The eigenfunctions of the impurity must follow the tetrahedral ($T_d$) site symmetry of the donor; thus $1s$, $2s$, and $2p_0$ states can be classified with the irreducible representations, namely, $A_1 \oplus E \oplus T_2$ as in Ref. [S2]. In the close vicinity of the donor site, the shift from the Coulomb potential with spherical symmetry lifts the six-fold degeneracy of $1s(A_1 \oplus E \oplus T_2)$ and separates it into $1s(A_1)$, $1s(E)$, and $1s(T_2)$. This decomposition is defined as valley-orbit coupling. As a result, $1s(A_1)$ drops down significantly from the value of the effective mass theory $\frac{m^*}{\epsilon^2} \cdot 13.6$ eV $\approx 35$ meV, while $1s(E)$ and $1s(T_2)$ remain close to it. When we consider the spin of the donor electron, the double group representations of $\overline{T}_d$, namely, the doublets $\Gamma_6$ and $\Gamma_7$ and the quadruplet $\Gamma_8$ as in Ref. [S1] are required. The $\Gamma_6$ spin-orbit coupling leads to splitting of the $1s(T_2 : \Gamma_4)$ into $1s(T_2 : \Gamma_7)$ and $1s(T_2 : \Gamma_8)$, since $\Gamma_4 \otimes \Gamma_6 = \Gamma_7 \oplus \Gamma_8$. Also for $\overline{T}_d$, $\Gamma_1 \otimes \Gamma_6 = \Gamma_6$ and $\Gamma_3 \otimes \Gamma_6 = \Gamma_8$, so $1s(A_1 : \Gamma_1)$ and $1s(E : \Gamma_3)$ turn into $1s(A_1 : \Gamma_6)$ and $1s(E : \Gamma_8)$, respectively. For excited states $2s$ and $2p_0$, the crystal field interaction has a weaker effect, hence the eigenstates of different irreducible representations $\Gamma_6 \oplus \Gamma_7 \oplus 2\Gamma_8$ are nearly degenerate. The above analysis can be visualized in Fig. III(c).

When we consider transitions between localized levels, the exact form of eigenstates is critical, while the exact splitting energy, which only appears directly in the Dirac-$\delta$ functions as shown later, is of much less importance. Therefore, it is unnecessary to resolve the energy levels beyond 2 meV resolution. The corresponding treat-

ment, where for example the $E$ and $T_2$-derived states are approximately degenerate, is shown by the rightmost side of Fig. III(c). We denote the states of $1s(A_1)$, $1s(E+T_2)$, $2s$-$2p_0$ and conduction band as the energy levels $\ell = 0, 1, 2$ and $c$. The corresponding energies $\epsilon$ are taken approximately as $-45$, $-34$, $-11$ and $0$ meV, respectively [S2]. The spin indices $\sigma$ are considered from the actual eigenstates, and then besides spin-up $\uparrow$ and spin-down $\downarrow$, the spin-mixed states $\cup$, $\cap$, $\sqcup$ and $\sqcap$ are also included, which will be expanded explicitly below.

To construct the eigenstates of the system, we start with the separate wavefunctions of conduction and localized electrons as follows [S3],

$$\psi_c^n(\mathbf{r}) = \exp(i\mathbf{k} \cdot \mathbf{r})\varphi_n(\mathbf{r}),$$
$$\phi_d(\mathbf{r}) = \sum_n \alpha_n F_n^\ell(\mathbf{r})\varphi_n(\mathbf{r}), \qquad (I)$$

where $n = \pm x, \pm y, \pm z$ are valley indices. $\mathbf{k}$ is the wavevector of the conduction electron, and $\varphi_n$ is the periodic Bloch function at the conduction band minimum of the $n$-valley. The detailed form of $\varphi_n$ can be obtained from the empirical pseudopotential method [S4]. The envelope functions $F_n^\ell$ for $1s$ and $2p_0$ states are respectively

$$F_n^{1s}(\mathbf{r}) = \frac{1}{\sqrt{\pi a_1^2 b_1}} \exp\left(-\sqrt{\frac{r^2 - x_n^2}{a_1^2} + \frac{x_n^2}{b_1^2}}\right),$$
$$F_n^{2p_0}(\mathbf{r}) = \frac{x_n}{\sqrt{\pi a_2^2 b_2^3}} \exp\left(-\sqrt{\frac{r^2 - x_n^2}{a_2^2} + \frac{x_n^2}{b_2^2}}\right). \qquad (II)$$

Here, the $x_n$ component of $\mathbf{r}$ is directed toward the axis of the $n$-valley. Parameters $a_1 \approx 2.5$ nm, $b_1 \approx 1.4$ nm, $a_2 \approx 4.2$ nm and $b_2 \approx 2.4$ nm are the corresponding Bohr radii. They are estimated by $a_\ell^2 = \hbar^2/(2m_t E_\ell)$

and $b_\ell^2 = \hbar^2/(2m_l E_\ell)$, where $E_\ell$ is the value of the energy level under effective-mass approximation. Effective masses $m_t \approx 0.19 m_0$ and $m_l \approx 0.92 m_0$ correspond to transverse and longitudinal orientations, respectively.

In the basis $(+x, -x, +y, -y, +z, -z)$, the valley-dependent coefficients $\alpha_n$ in Eq. (I) for different irreducible representations are

$$\begin{aligned}
&\tfrac{1}{\sqrt{6}}(1,1,1,1,1,1), \quad (A_1); \\
&\left.\begin{array}{l}\tfrac{1}{2}(1,1,0,0,-1,-1), \\ \tfrac{1}{2}(1,1,-1,-1,0,0),\end{array}\right\} (E); \\
&\left.\begin{array}{l}\tfrac{1}{\sqrt{2}}(1,-1,0,0,0,0), \\ \tfrac{1}{\sqrt{2}}(0,0,1,-1,0,0), \\ \tfrac{1}{\sqrt{2}}(0,0,0,0,1,-1)\end{array}\right\} (T_2).
\end{aligned} \quad \text{(III)}$$

We can denote the corresponding wavefunction of localized states as

$$\phi_0, \ (A_1); \quad \phi_+, \ \phi_-, \ (E); \quad \phi_x, \ \phi_y, \ \phi_z, \ (T_2). \quad \text{(IV)}$$

With the analysis at the beginning of this section, we can construct the exact states $\phi_d^{\ell\sigma}$ with spatial wavefunctions from Eq. (IV) and spinors $\chi^\uparrow$, $\chi^\downarrow$ for each $\ell$ and $\sigma$. Hence for the ground state,

$$\phi_d^{0\uparrow}: \quad \phi_0 \chi^\uparrow; \qquad \phi_d^{0\downarrow}: \quad \phi_0 \chi^\downarrow. \quad \text{(V)}$$

For $\ell = 1$, $\phi_d^{\ell\sigma}$ involves the highly spin-mixed states from $T_2$. We denote them as $\cup, \cap, \sqcup$ and $\sqcap$ [see Fig. III(c) and related discussion]. Then,

$$\phi_d^{1\uparrow}: \quad \tfrac{1}{\sqrt{2}}(\phi_x + i\phi_y)\chi^\uparrow, \ \phi_+\chi^\uparrow, \ \phi_-\chi^\uparrow; \quad \text{(VIa)}$$

$$\phi_d^{1\downarrow}: \quad \tfrac{1}{\sqrt{2}}(\phi_x - i\phi_y)\chi^\downarrow, \ \phi_+\chi^\downarrow, \ \phi_-\chi^\downarrow; \quad \text{(VIb)}$$

$$\begin{aligned}
\phi_d^{1\cup}: \quad &\tfrac{1}{\sqrt{6}}(\phi_x + i\phi_y)\chi^\downarrow + \sqrt{\tfrac{2}{3}}\phi_z\chi^\uparrow; \\
\phi_d^{1\cap}: \quad &\tfrac{1}{\sqrt{6}}(\phi_x - i\phi_y)\chi^\uparrow + \sqrt{\tfrac{2}{3}}\phi_z\chi^\downarrow;
\end{aligned} \quad \text{(VIc)}$$

$$\begin{aligned}
\phi_d^{1\sqcup}: \quad &\tfrac{1}{\sqrt{3}}(\phi_x + i\phi_y)\chi^\downarrow - \tfrac{1}{\sqrt{3}}\phi_z\chi^\uparrow; \\
\phi_d^{1\sqcap}: \quad &\tfrac{1}{\sqrt{3}}(\phi_x - i\phi_y)\chi^\uparrow - \tfrac{1}{\sqrt{3}}\phi_z\chi^\downarrow.
\end{aligned} \quad \text{(VId)}$$

$\phi_d^{\ell\sigma}$ for $\ell = 2$ simply includes all the combinations of Eq. (IV) and pure spinors, since we regard these states as degenerate. Meanwhile, the envelope function in Eq. (II) for $\ell = 2$ corresponds to $2p_0$ instead of $1s$ for $\ell = 0, 1$.

We can treat the conduction band bottom as extremely high-level donor state, and the corresponding Bohr radius can be considered as infinite. The wavefunction then turns to be the same as $\psi_c^n$ when $\mathbf{k} = 0$. We can simply use $\varphi_n$.

**3. Evaluation of rate coefficients.** All the coefficients denoted by $\omega$, $\eta$ and $\gamma$ in Eq. (1) of the main paper can be obtained from Fermi's golden rule

$$\Gamma_{i,f} = \frac{2\pi}{\hbar} |\langle \Psi_i | \mathcal{H}' | \Psi_f \rangle|^2 \delta[\epsilon(\Psi_f) - \epsilon(\Psi_i)], \quad \text{(VII)}$$

where $\Psi_i$ and $\Psi_f$ are the wavefunctions of the system for the initial and final states, respectively. $\mathcal{H}'$ is the perturbing Hamiltonian (not included in the system when obtaining $\Psi_i$ and $\Psi_f$) and the change of the system energy $\epsilon$ contributes in the Dirac $\delta$ function. In our case $\mathcal{H}'$ can be deformation potential or Coulomb interaction, for phonon-assisted process or exchange and impact effects.

With the explicit forms of wavefunctions shown in the previous section, we can evaluate the rate coefficients. For the phonon-assisted transitions, we can write the matrix element $\langle \Psi_i | \mathcal{H}' | \Psi_f \rangle$ in Eq. (VII) as

$$\langle \phi_d^{\ell\sigma} | \mathcal{H}_{\text{ep}} | \phi_d^{\ell'\sigma'} \rangle = \sum_{n,n'} \alpha_n^\ell \alpha_{n'}^{\ell'} \chi_n^\sigma \chi_{n'}^{\sigma'} \langle F_n^\ell \varphi_n | \mathcal{H}_{\text{ep}} | F_{n'}^{\ell'} \varphi_{n'} \rangle,$$

where $\alpha_n^\ell$ and $\chi_n^\sigma$ are the corresponding valley coefficient and spinor, respectively, in the donor-state wavefunction $\phi_d^{\ell\sigma}$. The smooth envelope function $F_n^\ell$ can be treated as constant on the scale of a unit cell. The matrix element of electron-phonon interaction between impurity states can then be expressed in terms of the same interaction between the conduction-band electrons, namely

$$\begin{aligned}
\langle F_n^\ell \varphi_n | \mathcal{H}_{\text{ep}} | F_{n'}^{\ell'} \varphi_{n'} \rangle = \qquad\qquad\qquad &\text{(VIII)} \\
\langle \psi_c^n | \mathcal{H}_{\text{ep}} | \psi_c^{n'} \rangle \int d^3\mathbf{r} F_n^{\ell*}(\mathbf{r}) F_{n'}^{\ell'}(\mathbf{r}) e^{i(\mathbf{k}-\mathbf{k}'-\mathbf{g}-\mathbf{q})\cdot\mathbf{r}}, &
\end{aligned}$$

where the extra phase comes from the detailed analysis of the electron-phonon interaction $\mathcal{H}_{\text{ep}}$ and the Bloch wave $\varphi_n$ [S5]. $\mathbf{g}$ is the reciprocal lattice; $\mathbf{q}$ is the phonon wavevector; $\mathbf{k}$ and $\mathbf{k}'$ are the corresponding wavevectors of initial and final states. We notice that coupling via intervalley phonons dominates [S6].

For the conduction band, we have [S7]

$$\begin{aligned}
\Gamma_{n,n'} &= \frac{2\pi}{\hbar} \left| \langle \psi_c^n | \mathcal{H}_{\text{ep}} | \psi_c^{n'} \rangle \right|^2 \left(N_\mathbf{q} + \tfrac{1}{2} \mp \tfrac{1}{2}\right) \delta(\Delta\epsilon \mp \hbar\Omega_\mathbf{q}) \\
&= \frac{\pi(D_t k)^2}{\rho V \Omega_\mathbf{q}} \left(N_\mathbf{q} + \tfrac{1}{2} \mp \tfrac{1}{2}\right) \delta(\Delta\epsilon \mp \hbar\Omega_\mathbf{q}), \quad \text{(IX)}
\end{aligned}$$

where $\Delta\epsilon = \epsilon(\mathbf{k}') - \epsilon(\mathbf{k})$, and $D_t k$ is the intervalley constant of deformation potential. $\rho$ and $V$ are the crystal density and volume. $N_\mathbf{q}$ and $\Omega_\mathbf{q}$ are the phonon distribution and frequency. The sign $-$ and $+$ corresponds to phonon absorption and emission, respectively. Hence we can write for the donor states that

$$\begin{aligned}
\omega_{\ell\sigma,\ell'\sigma'} = &\sum_{\mathbf{q},n,n'} \int \frac{d^3\mathbf{k}'}{(2\pi)^3} \frac{\pi(D_t k)^2}{\rho\Omega_\mathbf{q}} \\
&\times \left[\int d^3\mathbf{r} F_n^\ell(\mathbf{r}) F_{n'}^{\ell'*}(\mathbf{r}) e^{i(\mathbf{k}-\mathbf{k}'-\mathbf{g}-\mathbf{q})\cdot\mathbf{r}}\right]^2 \quad \text{(X)} \\
&\times (N_\mathbf{q} + \tfrac{1}{2} \mp \tfrac{1}{2}) \delta[\epsilon_{\ell'} - \epsilon_\ell \mp \hbar\Omega_\mathbf{q}] \alpha_n^\ell \alpha_{n'}^{\ell'} \chi_n^\sigma \chi_{n'}^{\sigma'},
\end{aligned}$$

which can be readily evaluated [S6]. We take $D_t k$ as 0.8 and 0.3 eV·Å$^{-1}$ for the dominant LA-$g$ and

TA-$f$ phonons respectively [S7], based on the phonon dispersion [S8]. From the Bose-Einstein distribution $N_q$ of phonons, we can approximate $\omega_{\ell\sigma,\ell'\sigma'} \approx \omega_{\ell'\sigma',\ell\sigma} \exp[-(\epsilon_{\ell'} - \epsilon_\ell)/(k_B T)]$ for $\ell' > \ell$ at low temperatures [S9]. Without $\sigma$ and $\sigma'$, namely omitting the $\chi_n^\sigma \chi_{n'}^{\sigma'}$ term in Eq. (X), we obtain $\omega_{1,0} \approx 4 \times 10^{11}$ s$^{-1}$, $\omega_{2,1} \approx 1 \times 10^9$ s$^{-1}$, $\omega_{c,2} \approx 3 \times 10^{10}$ s$^{-1}$. We notice the relatively slow transition between levels $\ell = 1$ and $\ell = 2$, which leads to the long-lived states claimed in favor of terahertz laser action [S10]. For the evaluation of $\omega_{2,c}$, we can approximately take $\epsilon = -6$ meV ($2p_\pm$ states) instead of $\epsilon = 0$ meV due to the continuum formed by highly excited localized states which extend to the conduction band [S2].

So far, single-phonon interactions do not directly couple the states on the same level. For $\omega_{\ell\sigma,\ell'\sigma'}$ when $\ell = \ell'$ we need to consider two-phonon processes with virtual states. Since it is a higher order term, we only consider it for $\ell = 0$ since it has the dominant electron density at low temperatures [S11]. The virtual states of $\ell' = 1$ are considered. From Eq. (VII) we can write [S12]

$$\omega_{0\uparrow,0\downarrow} = \frac{2\pi}{\hbar} \int \frac{d^3\mathbf{q}'}{(2\pi)^3} \int \frac{d^3\mathbf{q}}{(2\pi)^3}$$
$$\times \left| \sum_{\sigma'} \frac{\langle \phi_d^{0\downarrow} | \mathcal{H}_{\rm ep} | \phi_d^{1\sigma'} \rangle \langle \phi_d^{1\sigma'} | \mathcal{H}_{\rm ep} | \phi_d^{0\uparrow} \rangle}{\epsilon_1 - \epsilon_0} \right|^2 \quad \text{(XI)}$$
$$\times N_{\mathbf{q}}(N_{\mathbf{q}'} + 1) \delta(\hbar\Omega_{\mathbf{q}'} - \hbar\Omega_{\mathbf{q}}).$$

This term corresponds to the transition of spin-up ground state to the spin-down ground state via the virtual transitions to and from the spin-mixed first-excited states with phonon absorption and emission, respectively. This term is generally known as Casner-Orbach spin relaxation of the donor states [S12]. The matrix element can be evaluated similarly, or taken as $\omega_{0\uparrow,0\downarrow} = \omega_{\rm CO} \exp[-(\epsilon_1 - \epsilon_0)/(k_B T)]$ where $\omega_{\rm CO} \approx 2 \times 10^9$ s$^{-1}$ [S13]. $\omega_{0\downarrow,0\uparrow} = \omega_{0\uparrow,0\downarrow}$ can be obtained from symmetry (without magnetic field).

For the exchange coefficient $\eta$, we have [S14]

$$\eta_{c,\ell} = \frac{2\pi}{\hbar} \sum_{n,n',\ell'} \int \frac{d^3\mathbf{k}'}{(2\pi)^3} \left| \langle \psi_c^n, \phi_d^\ell | \mathcal{H}_{\rm C} | \phi_d^{\ell'}, \psi_c^{n'} \rangle \right|^2$$
$$\times \delta[\epsilon(\mathbf{k}') - \epsilon(\mathbf{k}) + (\epsilon_{\ell'} - \epsilon_\ell)], \quad \text{(XII)}$$

where $\mathbf{k}$ and $\epsilon(\mathbf{k})$ is the wavevector and kinetic energy of the conduction electrons, respectively. The matrix element for exchange scattering is

$$\langle \psi_c^n, \phi_d^\ell | \mathcal{H}_{\rm C} | \phi_d^{\ell'}, \psi_c^{n'} \rangle = \frac{e^2}{4\pi\varepsilon_0\kappa_s} \int d^3\mathbf{r}_1 d^3\mathbf{r}_2 \quad \text{(XIII)}$$
$$\times \psi_c^n(\mathbf{r}_1) \phi_d^\ell(\mathbf{r}_2) \phi_d^{\ell'*}(\mathbf{r}_1) \psi_c^{n'*}(\mathbf{r}_2) \left[ \frac{1}{|\mathbf{r}_1|} - \frac{1}{|\mathbf{r}_1 - \mathbf{r}_2|} \right].$$

This expression is merely the Coulomb interaction as a perturbation on the potential defining the initial state. The lack of a term $1/|\mathbf{r}_2|$ inside the square bracket comes from the fact that such Coulomb interaction is already accounted for in the initial wavefunction. Eqs. (XII)-(XIII) can be evaluated numerically [S15].

The coefficient $\gamma$ for the impact effects can be evaluated similarly to Eqs. (XII)-(XIII) as

$$\gamma_{\ell\sigma,\ell'\sigma'}^{\sigma''} = \frac{2\pi}{\hbar} \sum_{\sigma''',\ell'}^{n'',n'''} \int \frac{d^3\mathbf{k}'}{(2\pi)^3} \left| \langle \Psi_{n''\sigma'',\ell\sigma} | \mathcal{H}_{\rm C} | \Psi'_{n'''\sigma''',\ell'\sigma'} \rangle \right|^2$$
$$\times \delta[\epsilon(\mathbf{k}') - \epsilon(\mathbf{k}) + (\epsilon_{\ell'} - \epsilon_\ell)], \quad \text{(XIV)}$$

where the matrix element follows the form of Eq. (XIII) as

$$\langle \Psi_{n''\sigma'',\ell\sigma} | \mathcal{H}_{\rm C} | \Psi_{n'''\sigma''',\ell'\sigma'} \rangle = \frac{e^2}{4\pi\varepsilon_0\kappa_s} \int d^3\mathbf{r}_1 d^3\mathbf{r}_2 \quad \text{(XV)}$$
$$\times \Psi_{n''\sigma'',\ell\sigma}(\mathbf{r}_1, \mathbf{r}_2) \Psi_{n'''\sigma''',\ell'\sigma'}(\mathbf{r}_1, \mathbf{r}_2) \left[ \frac{1}{|\mathbf{r}_1|} - \frac{1}{|\mathbf{r}_1 - \mathbf{r}_2|} \right],$$

and the wavefunction of the system is constructed antisymmetrically

$$\Psi_{n'\sigma',\ell\sigma}(\mathbf{r}_1, \mathbf{r}_2) = \frac{1}{\sqrt{2}} \left[ \psi_c^{n'\sigma'}(\mathbf{r}_1) \phi_d^{\ell\sigma}(\mathbf{r}_2) - \psi_c^{n'\sigma'}(\mathbf{r}_2) \phi_d^{\ell\sigma}(\mathbf{r}_1) \right]$$

to guarantee the Pauli exclusion principle. This treatment is a generalization of the traditional problem for the scattering with singlet or triplet spin state [S16]. It can apply to the two-electron system when the spatial wavefunction and the spinor are not directly separable, as in the case when spin-mixed states of $T_2$ symmetry are involved. The classic results including spin asymmetry [S17] can be recovered in this formalism.

The spinors in Eqs. (XII)-(XV) contribute only to the summation of coefficients, while the spatial wavefunctions need to be integrated explicitly. Direct numerical calculations are rather time-consuming, so some simplification at the beginning is desirable. Here we show an example. Consider

$$M_{\rm ex} = \frac{e^2}{4\pi^2\varepsilon_0\kappa_s a^{*3}} \int d^3\mathbf{r}_1 d^3\mathbf{r}_2 \exp[\,i\,(\mathbf{k}\cdot\mathbf{r}_1 - \mathbf{k}'\cdot\mathbf{r}_2)]$$
$$\times \exp\left(-\frac{|\mathbf{r}_1| + |\mathbf{r}_2|}{a^*}\right) \left[ \frac{1}{|\mathbf{r}_1|} - \frac{1}{|\mathbf{r}_1 - \mathbf{r}_2|} \right], \quad \text{(XVI)}$$

which appears in the spatial part of Eq. (XIII) or Eq. (XV) after the Bloch waves $\varphi_n$ are first integrated and lead approximately to a Kronecker-$\delta$ function. Spherical symmetry is also approximated for the envelope function $F_n^\ell$ where $a^* = \sqrt[3]{(a^2 b)}$.

The two terms in the square bracket of Eq. (XVI) can be evaluated separately as

$$M_{\rm ex}(\mathbf{k}, \mathbf{k}') = \frac{e^2}{\varepsilon_0\kappa_s} \frac{8a^{*2}}{[1 + (a^*\mathbf{k})^2][1 + (a^*\mathbf{k}')^2]^2} - M_0,$$

where

$$M_0 = \frac{e^2}{\varepsilon_0 \kappa_s} \frac{\alpha'^2}{4\pi^2} \int d^3\boldsymbol{\rho}_1 d^3\boldsymbol{\rho}_2 \exp[i\,a^*(\mathbf{k}\cdot\boldsymbol{\rho}_1 - \mathbf{k}'\cdot\boldsymbol{\rho}_2)]$$
$$\times \exp(-|\boldsymbol{\rho}_1| + |\boldsymbol{\rho}_2|) \frac{1}{|\boldsymbol{\rho}_1 - \boldsymbol{\rho}_2|} \,. \quad \text{(XVII)}$$

Here, $\boldsymbol{\rho}_1 = \mathbf{r}_1/a^*$ and $\boldsymbol{\rho}_2 = \mathbf{r}_2/a^*$. From Fourier transformation

$$\frac{1}{|\boldsymbol{\rho}_1 - \boldsymbol{\rho}_2|} = 4\pi \int \frac{d^3\mathbf{q}}{(2\pi)^3} \frac{\exp[i\,\mathbf{q}\cdot(\boldsymbol{\rho}_1 - \boldsymbol{\rho}_2)]}{\mathbf{q}^2} \,,$$

so that we can reduce Eq. (XVII) to

$$M_0 = \frac{e^2}{\varepsilon_0 \kappa_s} \frac{8a^{*2}}{\pi^2} \int \frac{d^3\mathbf{q}}{\mathbf{q}^2[1+(a^*\mathbf{k}+\mathbf{q})^2]^2[1+(a^*\mathbf{k}'+\mathbf{q})^2]^2} \,,$$

where the integration is over all $\mathbf{q}$ the $k$-space. In spherical coordinates, we denote $\mathbf{q} = (q, \vartheta, \varphi)$ and integrate over $q$ by contour integration [S15]. Thus,

$$M_0 = \frac{e^2}{\varepsilon_0 \kappa_s} 8a^{*2} \int_\Omega \frac{\sin\vartheta d\vartheta d\varphi}{4\pi} \frac{1}{(d_i d_f D)^2}$$
$$\times \left[ \frac{d_i^3 + d_f^3}{d_i d_f} + \frac{4(d_i + d_f)^3}{D} \right] \,, \quad \text{(XVIII)}$$

where the integration is over all solid angles $\Omega$ and

$$d_i^2 = 1 + (a^*\mathbf{k})^2 \left(1 - \cos^2\langle\mathbf{k},\mathbf{q}\rangle\right) \,,$$
$$d_f^2 = 1 + (a^*\mathbf{k}')^2 \left(1 - \cos^2\langle\mathbf{k}',\mathbf{q}\rangle\right) \,, \quad \text{(XIX)}$$
$$D = (d_i + d_f)^2 + a^{*2} \left(|\mathbf{k}|\cos\langle\mathbf{k},\mathbf{q}\rangle - |\mathbf{k}'|\cos\langle\mathbf{k}',\mathbf{q}\rangle\right)^2 \,.$$

For a general scattering angle $\theta = \langle\mathbf{k}, \mathbf{k}'\rangle$ we can numerically evaluate Eq. (XVIII). Taking the polar angle $\vartheta$ of $\mathbf{q}$ to be measured from the axis along $\mathbf{k}$ and the azimuthal angle $\varphi$ from the plane constructed by $\mathbf{k}$ and $\mathbf{k}'$, we have

$$\cos\langle\mathbf{k},\mathbf{q}\rangle = \cos\vartheta, \quad \cos\langle\mathbf{k}',\mathbf{q}\rangle = \cos\theta\cos\vartheta + \sin\theta\sin\vartheta\cos\varphi$$

for Eqs. (XVIII)-(XIX). We notice that the matrix element $M$ depends only on $k = |\mathbf{k}|$, $k' = |\mathbf{k}'|$ and $\theta = \langle\mathbf{k}, \mathbf{k}'\rangle$. Then we can write $M_{\text{ex}}(\mathbf{k}, \mathbf{k}') = M_{\text{ex}}(k, k', \theta)$, and it can be readily evaluated numerically.

We notice that the $\omega$ coefficient does not depend on the distribution of conduction electrons, yet the $\eta$ and $\gamma$ coefficients do. For the master equations of the whole system, we need to integrate over all conduction electrons, which will undergo drift and heating in the applied electric field. We can expect strong electric field dependence of $\eta$ and $\gamma$, which is consistent with the analysis in the main paper. We use Monte Carlo simulation to generate the distribution of conduction electrons [S7, S18]. In this way, all the rate coefficients in Eq. (1) of the main paper can be evaluated numerically for any value of electric field and temperature.

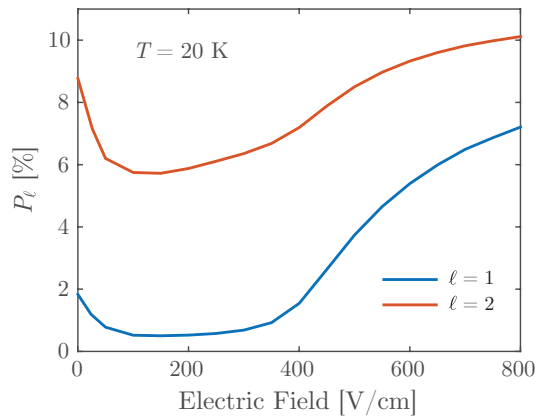

FIG. IV. Dynamic polarization of bound electrons in the $1s(E)/1s(T_2)$ ($\ell = 1$) and $2s/2p_0$ ($\ell = 2$) states at $T = 20$ K, with injected conduction electron polarization of 11.5%.

**4. Solution to master equations.** Although we are only interested in the steady-state solution, the system of rate equations in Eq. (1) of the main paper (without an additional constraint of local charge neutrality) must be solved explicitly in time, because they are not independent when $\partial n_{\ell\sigma}/\partial t = 0$. We use initial conditions that the impurities are completely depleted, i.e. at $t = 0$, $\sum_\sigma n_{\ell\sigma} = 0$ for $\ell \neq c$, and allow the densities to evolve in time. The structure of the equations is such that total charge is automatically conserved during time evolution; the density of conduction electrons used to generate Fig. 3(b)-(c) of the main text is obtained at convergence.

**5. Dynamic polarization of bound electrons.** To justify our assertion that the model presented here may be relevant for simulation of THz emission from impurity level transitions, we present Fig. IV. A substantial spin polarization of the electrons occupying the $2s$ and $2p_0$ excited states ($\ell = 2$) even in the intermediate-field impact excitation regime indicates that the dynamic spin polarization of localized electrons can be controlled by external injection from a ferromagnetic source; appropriate selection rules for radiative relaxation transitions from these spin-orbit-split excited states to the $1s$ levels may provide a means for electrical control over electromagnetic polarization.

---



\end{document}